\documentclass[runningheads,a4paper]{llncs}

\usepackage{amssymb}
\usepackage{graphicx}
\usepackage{bm}

\usepackage{url}
\urldef{\mailsa}\path|{grohg, schams}@in.tum.de|    
\newcommand{\keywords}[1]{\par\addvspace\baselineskip
\noindent\keywordname\enspace\ignorespaces#1}

\begin{document}

\mainmatter  

\title{Contributive Social Capital Extraction \\
From Different Types of Online Data Sources} 

\titlerunning{Social Capital and Reputation in Social Networks}

\author{Sebastian Schams \and Georg Groh}
\authorrunning{Social Capital and Reputation in Social Networks}

\institute{Technical University of Munich, Germany, May 2018}

\toctitle{Social Capital}
\tocauthor{in Social Networks}
\maketitle

\begin{abstract}
It is a recurring problem of online communication that the properties of unknown people are hard to assess. This leads to various issues like the rise of so called `fake news' from untrustworthy sources, or simply not being able to assess the value of a membership in a social network, especially regarding information and data access.

In sociology the sum of (social) resources available to a person through their social network is often described as social capital. 
In this article, we look at social capital from a different angle. Instead of evaluating the advantage that people have because of their membership in a certain group, we investigate various ways to infer the social capital a person adds or may add to the network, their contributive social capital (CSC). As there is no consensus in the literature on what the social capital of a person exactly consists of, we look at various related properties: expertise, reputation, trustworthiness, and influence.

The analysis of these features is investigated for five different sources of online data: microblogging (e.g., Twitter), social networking platforms (e.g., Facebook), direct communication (e.g., email), scientometrics, and threaded discussion boards (e.g., Reddit). In each field we discuss recent publications and put a focus on the data sources used, the algorithms implemented, and the performance evaluation. The findings are compared and set in context to contributive social capital extraction.

The analysis algorithms are based on individual features (e.g., followers on Twitter), ratios thereof, or a person's centrality measures (e.g., PageRank). The machine learning approaches, such as straightforward classifiers (e.g., support vector machines) use ground truths that are connected to social capital. The discussion of these methods is intended to facilitate research on the topic by identifying relevant data sources and the best suited algorithms, and by providing tested methods for the evaluation of findings.

\end{abstract}

\section{Introduction}

\subsection{Motivation} \label{se:motivation}
In the last years, many interactions between people and between people and companies moved from the offline world into the realms of the more or less anonymous online world. People have been dealing with assessing each other in face-to-face interactions for a long time. 
It is generally a clear advantage if one can infer more properties of other people like trustworthiness, reputation, expertise, or influence. This helps to understand where the competence of another individual lies, see their motives, and make fact-based decisions based on the impressions. 
If the counterpart is an unknown online individual one is faced with two differences. For one there is much less direct information available (e.g., social signals such as voice prosody or facial expressions) and secondly, there is a variety of new indicators (in this paper called intrinsic features) that provide indirect information about the user (e.g., likes on Facebook or followers on Twitter). Dealing with this different information setting is complicated because the intrinsic features by themselves may be misleading and can be manipulated.

Internet communities have introduced feedback systems that are founded on interactions between users and are supposed to create transparency. The first step is usually to ask new users to register, so that they are bound to a public profile that allows others to form an opinion based on previous activities or other users' feedback. How this is implemented differs from platform to platform, depending on the main purpose of the community. Amazon allows users to rate sellers on a scale from one to five stars \cite{Bhattacharjee2005}. Twitter has retweeting and following \cite{Anger2011}, which provides information regarding the popularity of users, and Facebook allows you to ``like'' company profiles or other users' posts \cite{Smith2012} so one can see which posts or companies have the most support.

However, all systems do have shortcomings. Sometimes the limitations are a result of the community's focus on other priorities. Facebook, for example, is interested in maximizing the time members spend on the website, rather than discouraging people to post by visualizing their limited competencies in certain areas. In other cases, and especially regarding social capital, these measures do not go far enough, because they are not necessarily directly correlated with social capital.

Therefore, the true CSC of a person stays hidden, the expertise of a poster of social media comments cannot be fully assessed, and the trustworthiness or reputation of someone is as hard to infer as their influence --- at least on the basis of feedback systems alone.

In this paper we discuss social capital for an individual and how it relates to other properties. Then we present and compare various methods to extract these properties from online data sources in ways that yield better results than the intrinsic methods and relate these findings to the task of identifying individual CSC scores.


\subsection{Outline of the paper}
In section \ref{se:definitions} we discuss how (contributive) social capital is defined in general and what definitions are most relevant for our context. It is then set into context with reputation, trust, expertise, and social influence. Section \ref{se:intrinsicimplementation} discusses the data sources that are most relevant for online interactions and what intrinsic measures they offer to infer contributive social capital and its aspects. In section \ref{se:extractionmethods} we present and compare selected methods from previous work that have been successfully used to extract reputation, trust, expertise, or social influence from the different data sources. The findings are discussed in the context of contributive social capital because the current research on direct CSC inference from the discussed sources is limited. Finally, the findings are summarized in section \ref{se:summary}.

\section{Relevant terms and their overlap}\label{se:definitions}
In sociology the term `social capital' is often used to describe a property of social networks as a whole. Exemplary for this is Putnam's definition from 1995, who describes social capital as ``features of social organization such as networks, norms, and social trust that facilitate coordination and cooperation for mutual benefit'' \cite{Putnam95}. Social capital can also be analyzed on an individual level. This is usually done by analyzing the availability of social resources a person has through their relations with others within the network (see also the definitions by \cite{lin2002}, \cite{Recuero2011}, and \cite{Phulari2010} discussed in section \ref{su:definitions}). In this paper, we want to focus on social capital from a different point of view. Instead of looking at a network from the point of view of an individual, we treat the social capital of a person as the social capital this individual adds to the network --- their \textbf{contributive social capital (CSC)}. To measure and extract this type of social capital from online interactions, we first need to define the dimensions of contributive social capital.

As CSC is used to describe interactions between people, all attributes of a person that are of relevance for human interactions are important. In this paper we focus on several factors, for which analysis mechanisms have been investigated in the past:
\begin{itemize}
	\item Factual competence (e.g., the knowledge, experience, and expertise a person brings to the network),
	\item Trust and reputation (e.g., the degree to which others can believe that information provided by this individual is true and that her actions are benevolent),
	\item Social influence (e.g., as a result of someone being valuable to their social network).
\end{itemize}

These characteristics represent an important part of anyone's online persona. In the following we have a closer look at these terms and set them in relation to CSC to motivate this approach.

\subsection{Definitions}\label{su:definitions}

\subsubsection{Reputation.}
Hoffman et al. define reputation as ``the opinion of the public towards a person, a group of people, an organization, or a resource'' \cite{Hoffman2009}. This is consistent with other definitions (e.g., compare J\o{}sang et al. \cite{JAsang2007} who quote the Concise Oxford dictionary: ``Reputation is what is generally said or believed about a person's or thing's character or standing''). In the context of social networks this means that the reputation of a participant is built with information that is available to the social network. Reputation can therefore be defined as the sum of all subjective judgments of actors in a network regarding a single individual. From these definitions it becomes apparent that reputation is a subjective element, a degree of belief to which others intersubjectively believe in another person's competence in a certain field.

\subsubsection{Trust.} 
An even more individually subjective construct is trust. People can choose to trust or distrust others without considering their respective public reputation. Diego Gambetta's definition from 1988 is often used throughout literature: ``trust (or, symmetrically, distrust) is a particular level of the subjective probability with which an agent assesses that another agent or group of agents will perform a particular action'' \cite{Gambetta88}. J\o{}sang et al. \cite{JAsang2007} contrast this definition with a quote from Mcknight et al.\cite{Mcknight1996} on what they call ``Trusting Intention'': ``the extent to which one party is willing to depend on the other party in a given situation with a feeling of relative security, even though negative consequences are possible''. 
This underlines that trust is usually an assessment of an individual, whereas the reputation is the average of all subjective judgments of actors in a network. Similar to reputation, trust is usually topic-focused and situation-dependent. The average person would, for example, trust a math teacher more with answering a mathematical problem than a person chosen at random. This aspect of trust is relevant for online interactions, especially when users primarily care about a single topic. A seller on an online auctioning site is, for instance, mainly interested in the buyer's willingness to pay. On the other hand, the reader of a threaded discussion board wants to know how competent the author of certain posts is in the topic and whether she can believe the recommendations.

\subsubsection{Expertise.} 
If we move from the example of buyer-seller networks to content-oriented internet platforms (e.g., threaded discussion boards or expert exchanges), the expertise of an information provider gets increasingly important. According to Bozzon et al., expertise is closely related to someone's knowledge in a defined domain. They describe identifying experts within a social network as the task of ``ranking the members of a social group according to the level of knowledge that they have about a given topic'' \cite{Bozzon2013a}. After the ranking, the top \textit{k experts} may be chosen and selected to answer questions or fulfill tasks. Therefore, their expertise can be regarded as the ability to answer questions on certain topics with the help of their knowledge or experience.

Expertise is in general verifiable based on facts, whereas reputation and trust are more subjective and less specific judgements. 

\subsubsection{Social influence.} 
A widely studied research subject is \textbf{social influence}. According to J. Sun, ``social influence refers to the behavioral change of individuals affected by others in a network'' \cite{Sun2011a}. With the emergence of online social networks that document a wide range of user interactions, it became possible to study influence regarding a wide range of aspects and fields. Anger et al. conclude in \cite{Anger2011} that only a small percentage of all users on Twitter are what they call ``influencers.'' These people publish content that is in turn read and reposted by many followers. Identifying these people is of a high interest to companies for marketing and directed advertisements \cite{Sun2011a}.


\subsubsection{(Contributive) social capital.} 
A term that includes elements of all the aforementioned areas and focuses on the relationships and network of a person is \textbf{social capital}. It can be understood as a measure to depict the civic participation, political engagement, and life satisfaction of a person (as in Valenzuela et al. \cite{Valenzuela2009}). Another popular definition is more applicable in the context of this paper: ``Social capital is a broad concept, usually focused on the values obtained by being part of a social network and thus, referred to as the sum of `social resources' '' \cite{lin2002}. Lin et al. therefore describe the social capital of a person as something that is obtained by participating in a network, adopting the social norms and accessing the social resources. Adding to that, Recuero et al. conclude from several authors that ``the most common form of social capital is information access'' \cite{Recuero2011}. This illustrates the two facets of social capital as something based on subjective judgments by others on the one hand as well as a usable resource on the other hand. Another definition was published by Phulari et al. in 2010: ``Social capital is an elastic term. It broadly refers to the resources accumulated through the relationships among people.'' \cite{Phulari2010}. Recuero et al., Lin et al., and Phulari et al. all underline the value-add a person experiences by being part of a network, e.g., in the form of shared norms and information access, as well as by providing resources. 

As stated in the beginning of this section, this paper focuses on the social capital of an individual person for their network, which we call \textbf{contributive social capital}, and how it can be extracted from online data sources. To the best of our knowledge there is no tested and verified approach to directly accomplish this task. Therefore, we investigate publications that deal with the identification of related attributes -- reputation, trust, expertise, influence -- and set them into context with CSC, which we propose as umbrella term to cover all relevant aspects of the four terms. For that, we first have to look at the relations between these terms and contributive social capital. 







\subsection{Overlap between contributive social capital and reputation, trust, expertise, influence}\label{su:overlap}
In the last section we have seen that CSC, like the other terms, is generally linked to and influences a person's value within a network. This value is primarily influenced by that person's relationships to others and by the information and resource access that person has.
Therefore, all attributes of a person that facilitate social interactions, foster relationships, and promote information access can be regarded as aspects of (contributive) social capital.

\textbf{Reputation} increases a person's CSC, as other people are more likely to connect and interact with, as well as provide information to a reputable person. This was investigated by Resnick et al. \cite{Resnick2000} on online buyer-seller networks. On eBay they observed that ``sellers with stellar reputations may enjoy an extra premium on their services --- a premium that users may be willing to pay for the security and the comfort of high quality services.'' If customers are willing to pay a premium price for a perceived better reputation, it is reasonable to assume that a higher reputation also correlates to non-monetary benefits.
A similar reasoning can be applied to \textbf{trust}. Valenzuela et al. argue based on Putnam's research, ``social trust facilitates associative behavior, fosters a strong civil society, and makes political institutions and officials more responsive [...]'' \cite{Valenzuela2009}. This can be understood as an increase of CSC.

The \textbf{expertise} of a person directly correlates with the information and resource access aspect of CSC (see e.g., \cite{Recuero2011}). It increases the amount of information potentially made available to the network by a person. People who are close to this individual can now access this information (e.g., by discussions). This leads to a similar effect we have seen for reputation: The overall increase of social capital in the network can be attributed to the individual CSC of the person.

There is also a significant overlap between a user's \textbf{influence} and his CSC. Influence usually stems from several factors. Among those are the average novelty of a user's contributions\cite{keller2003influentials}, their eloquence \cite{keller2003influentials}, and the quality of the content they produce \cite{Carmel2012}. These factors act as local multipliers on the CSC of a person and increase its basic constituting components directly. \par

Measuring the social capital of an individual in online interactions can therefore be correlated with these attributes. 


\section{Intrinsic implementation of CSC in online data sources}\label{se:intrinsicimplementation}

In this section we investigate how CSC (including reputation, trust, expertise, and influence) is formalized on existing platforms. This section is structured according to five major data sources that are relevant for online interactions:

\begin{itemize}
\item Microblogging (e.g., Twitter\footnote{https://twitter.com/ (retrieved 2017-08-04)})
\item Social networking platforms (e.g., Facebook\footnote{https://www.facebook.com (retrieved 2017-08-04)})
\item Direct conversation (e.g., email)
\item Citation networks (e.g., scientific citation networks)
\item Threaded discussions (e.g., Stack overflow\footnote{https://stackoverflow.com (retrieved 2017-08-04)}, Reddit\footnote{https://www.reddit.com/ (retrieved 2017-08-04)})
\end{itemize} 

We want to give an overview of possibilities to use these data sources for the identification of contributive social capital of network participants. For that, it is important to understand the general functionality of each platform and what intrinsic metrics can be used to approximate CSC. In the following, each data source is discussed by means of an example.

\subsubsection{Microblogging.} Twitter is an example of a \textbf{microblogging} \cite{Java2007} service with over 300 million active users per month and the self-declared mission ``to give everyone the power to create and share ideas and information instantly, without barriers''\footnote{https://about.Twitter.com/company (retrieved 2016-06-30)}.
Twitter messages have a maximum length of 280 characters and can contain text, URLs, pictures, mentions (references to other users with the symbol @ and their name), hashtags (i.e., a type of label that is usually connected to the post's context or a current event), and locations. Mentions can increase the popularity of the mentioned user, however, this popularity is not explicitly stated. Hashtags can increase the visibility of tweets because interested users can search for certain hashtags. Participants can follow others to receive their tweets directly \cite{Weng2010}, which is like a subscription. The connection between followers is often based on reciprocity, which was shown by Weng et al., who found that ``80.5\% of users have 80\% of users they are following follow them back'' \cite{Weng2010}. If a user agrees with a post it is possible to ``retweet'' the full comment or to quote the text and add a personal statement. Both acts increase the original post's visibility as it is made accessible to a larger crowd (namely the followers of the retweeter).
It has been shown by Hofer et al. that there is a correlation between Twitter usage and social capital because Twitter allows members to stay in touch with many people \cite{Hofer2013}. In their study participants, who spent more time on Twitter, also reported a higher perceived online bridging social capital. Additionally, Hofer et al. investigated how the numbers of followers and followees correlate to a user's online social capital and found a positive correlation as well: People with more followees/followers usually have a higher bridging/bonding social capital \cite{Hofer2013}. As the time spent on Twitter correlates with a user's social capital, it is reasonable to assume that the other metrics that signal a user's engagement correlate with their (contributive) social capital as well. One of these indicators could be the number of posts a user published. Indicators that include the perspective of other users might be even more useful: for example, the number of retweets users generate or the times they are mentioned in other posts. Users can, however, be mentioned by others who disagree with them fully. This is an interaction that increases the number of times they are mentioned, but it does not necessarily correlate with a high CSC. Analysis methods that implement and go beyond these single performance indicators are presented in section \ref{se:extractionmethods}.

\subsubsection{Social networking platforms.} 
Social networking platforms are online services that allow their members to connect with each other. 
Facebook is the most popular social networking platform and has over 2 billion active users\footnote{https://de.newsroom.fb.com/news/2017/06/facebook-bedankt-sich-bei-2-milliarden-monatlich-aktiven-menschen/ (retrieved 2018-04-19)}. 
The number of friends, likes, received and posted messages can all be used as indicators for a user's importance for the network and can therefore be regarded as connected to this user's CSC. Burke et al. argue in \cite{Burke2011} that ``it is plausible that creating and consuming undirected messages, allowing users to keep in touch, will lead to increases in social capital''. 
Similar to the previous example of Twitter, the passive indicators (received likes, friends, comments, and responses to own posts) are more trustworthy indicators for a person's value within the network because they incorporate feedback from others and are therefore more complex to manipulate.


\subsubsection{Direct communication.} There are many ways to directly communicate with another person online. Microblogging and social networks often implement forms of direct communication by allowing users to send messages to each other. Even more straightforward direct conversation is the use of mobile messengers or email. A popular mobile messenger is WhatsApp\footnote{https://www.whatsapp.com/ (retrieved 2018-04-19)}. 
For the purpose of CSC identification email has certain advantages. The length of a message is restricted neither by the service provider, nor by the ease of input, which is less comfortable for long texts on a smartphone. Also, it can include personal as well as professional information and therefore a more complete portrayal of someone's CSC is possible. As this data is usually private, there are no built-in indicators to visualize social capital to the public. However, there are data sets available that allow scientists to investigate social capital and its components in this context (e.g., Enron email data\footnote{Cohen, W. Enron email data set, http://www- 2.cs.cmu.edu/~enron/ (retrieved 2018-04-19)}). Metrics that can be used for analysis are the number of outgoing and incoming emails as well as the message content (see section \ref{se:extractionmethods}).

\subsubsection{Scientometrics.} Scientific citation networks and co-citation networks are formed by publications that reference other papers. 
Kas et al. list several properties of citation networks in \cite{Kas2012}. Citation networks are directed, acyclic networks where the bulk of the network is static and only the leading edge is dynamic \cite{Kas2012}. J\o{}sang et al. \cite{JAsang2007} also mention another important property: there are only positive referrals. Therefore, it is not easily possible to sanction authors whose publications were highly cited in the beginning but later turned out to be controversial.

Compared to the previous networks, citation and co-citation networks evolve slowly and there is no real exchange of directed communication content between people. In the context of social capital extraction, however, the fact-based scientific nature offers distinct advantages. The content is reviewed before it is published, which guarantees a high level of quality. Therefore, performance metrics can be used to draw conclusions about the expertise of author and co-authors.

The direct performance metric in citation networks is the number of citations an author or paper receives by other published articles.


\subsubsection{Threaded discussion boards.} These platforms organize the discussions of their users by topics in threads. The website Reddit is a popular representative of such a community and one of the most frequented websites in the world\footnote{http://www.alexa.com/siteinfo/Reddit.com (retrieved 2016-07-01) shows a global rank of 27.}. Reddit is a social news website with discussion threads \cite{Weninger2013} and is organized along different `Subreddits' that focus on specific topics (e.g., news or the NBA). Users can either link other online content and briefly summarize it with a headline, or post thoughts of their own. To promote the best content, Reddit has a voting mechanism that allows registered users to up- or down-vote content. The resulting score (the difference between up- and down-votes a user receives) is called \textit{karma} and comes in two flavors, link karma and comment karma \cite{Bergstrom2011}. The first one is awarded for links to other websites (e.g., news articles) that the community regards as worth reading, and the latter is assigned to a user's comments for either its informative content, its originality, or other aspects, like humor.

Another news and threaded discussion board is Slashdot \cite{Lampe2007}. The content posted on Slashdot is often related to science, technology, or politics and summarized by the posters. The posts are organized along different topics, and users can comment on them to discuss with each other. The voting system is slightly different from Reddit. Only randomly assigned moderators can up- or downvote posts, not every registered user. 

\section{Extraction methods of social capital}\label{se:extractionmethods}

In the last section, five different data sources were presented and specific performance metrics which might be used for CSC extraction were discussed. These metrics are limited to their respective platforms and usually have the primary goal of increasing overall customer satisfaction. This section presents a selection of more complex analysis methods for social capital identification --- again split along the five data sources microblogging, social networking platforms, direct communication, scientometrics, and threaded discussion boards. These extraction methods focus on inferring CSC constituents like trust, reputation, expertise, or influence from the network. As discussed in sections \ref{su:overlap} and \ref{su:definitions}, CSC is used as an umbrella term of these properties. Consequently, methods that focus on one of these areas can be used as input or be extended to extract CSC as well. We aim at making their similarities apparent and the methods comparable by discussing the implications for CSC extraction.


\subsection{Microblogging}  \label{chap4twitter}
Twitter has been studied in a large number of research papers. However, there is no consensus on what exactly social capital is on Twitter and how it is measured. The main research focus often lies in social influence, especially identifying influential users. This is a question relevant not only for scientific research but also for the marketing departments of companies. Handing out free samples to the biggest influencers can lead to successful advertising because these influencers often share their positive experiences and can spread information very effectively throughout the network \cite{Rao2015}.
There are several commercial services that analyze influence on Twitter. A prominent example is the Klout index, which will be discussed in subsection \ref{socialnetworkingplatforms}. The nature of discussions on Twitter is very content-oriented \cite{Robles2011}. Therefore, it is reasonable to assume that influential people on Twitter also have a higher than average CSC. 
With the help of the intrinsic metrics provided by Twitter and described in section \ref{se:intrinsicimplementation}, one can easily define more detailed performance indicators.

\subsubsection{Measuring influence with performance ratios.}
In their 2011 paper Anger et al. \cite{Anger2011} described and analyzed several performance ratios that are relevant for measuring someone's influence. The goal of each of the ratios is slightly different and described in the following. The same abbreviations as in \cite{Anger2011} are used.
\begin{itemize}
\item The \textit{Follower/Following Ratio ($r_{f}$)} compares the number of users who have subscribed to the updates of user A with the number of users that user A is following \cite{Anger2011}. The larger this number is, the more people are following the user without a reciprocal relationship.
\item To detect how many of user A's tweets imply a reaction from the audience, the \textit{Retweet and Mention Ratio ($r_{RT}$)} can be determined. This is the number of tweets that are amplified or lead to a communicative action between user A and another user, divided by the total number of tweets of user A
\cite{Anger2011}. This is a direct measure of influence of one's tweets. The more reactions a post provokes, the more influence it has had on other network participants.
\item To assess how many different individual users interact with user A, the \textit{Interactor Ratio ($r_{i}$)} can be calculated. This is the number of individual users who retweet content or mention user A, divided by the total number of followers of user A. \cite{Anger2011}
\item The \textit{social networking potential} (SNP) was introduced to take the content focus of the Retweet and Mention Ratio as well as the relational focus of the Interactor Ratio into account. The arithmetic mean of $r_{RT}$ and $r_{i}$ forms this new metric. The SNP thereby combines many important features: number of followers, individual interactors, retweets, mentions, and tweets. \cite{Anger2011}
\end{itemize}

Anger et al. analyzed these ratios on a data set of Austrian Twitter users from 2011. Their findings, with regard to the four different ratios, are summarized in the following, and the advantages and shortcomings of each method are discussed.
\begin{itemize}
\item The number of followers is one of the most often used metrics for influence on Twitter. By itself this number is prone to error: a lot of people follow others in the pure hopes of being followed back --- which is often successful \cite{Anger2011}. Furthermore, there are several services that sell followers \cite{Anger2011}. In addition to that, the pure number of followers does not tell anything about their own activity level or social capital. All these issues can skew the effectiveness of the $r_{f}$. Anger et al. created a list of the top ten Austrian Twitter users, ranked by number of Austrian followers. The number one ranked person has a $r_{f}$ of 130.1, meaning that the number of followers is about 130 times larger than the number of people he is following. Most of the other users in the top ten list, however, have a single-digit score. One user in the top ten list has a ratio of only 1.0 --- he is following as many users as are following him. Considering that he is still among the user's with the most followers illustrates the shortcomings of the Follower/Following Ratio.
\item The Retweet and Mention Ratio is a measure of how involved the influencers are with other users. The top three Austrian Twitter users, identified with this measure, is different to the top three found with the $r_{f}$ metric. This illustrates the different nature of both ratios. \cite{Anger2011}
\item Ranking the Austrian Twitter users according to the Interactor Ratio shows yet another top three. The differences are not surprising, because this ratio focuses more on conversation-oriented interactions, as it is increased by every mention or retweet a person receives. \cite{Anger2011}
\item The SNP is of interest to us because it takes content ($r_{RT}$) as well as personal relations ($r_{i}$) into account and both are important for social capital, which is influenced by information access as well as relations. The resulting top three are all known from the top three lists of the previous ratios, which is due to the fact that the formula is the arithmetic mean of $r_{i}$ and $r_{RT}$. \cite{Anger2011}
\end{itemize} 

Especially the content-focused $r_{RT}$, as well as the SNP; and to a smaller extent, the $r_{i}$ are of interest for social capital research. It can be assumed that people score high on these ratios when their post quality is high ($r_{RT}$, SNP), which shows that they contribute to the network, or that they have good relations with others ($r_{I}$), which is also an indicator for value within the network. High values on these ratios are consequently indicators that the person has a with high contributive social capital.

\subsubsection{PageRank algorithm for social network analysis.}
Other work employs and extends Google's famous PageRank algorithm, which is a centrality measure \cite{Kosch2004}. Google uses this, among other methods, to rank websites that show up as search results. Wu et al. summarize the functionality of PageRank in \cite{Wu2007}: 
\begin{quotation}
``In essence, PageRank relies on the democratic nature of the Web by using its vast link structure as an indicator of an individual page's quality. It interprets a hyperlink from page x to page y as a vote, by page x, for page y. Additionally, PageRank looks at more than just the sheer number of votes or links that a page receives. It also analyzes the page that casts the vote. Votes cast by pages that are themselves important weigh more heavily and help to make other pages more ``important''.'' \cite{Wu2007}
\end{quotation}

A similar notion can be applied to users of social networks instead of web pages. In this case the nodes are users instead of websites and the edges are votes exchanged between users, or similar metrics, instead of hyperlinks. Instead of the importance of a website, one can then analyze the contributive social capital of a user within the network. An early attempt to use a PageRank-like algorithm to research Twitter was the introduction of TunkRank by Daniel Tunke\footnote{http://thenoisychannel.com/2009/01/13/a-Twitter-analog-to-pagerank/ (retrieved 2016-08-08)}. It is a recursive method to calculate the influence of a person X based on the influence of the followers and is defined as follows:
	\begin{equation}
Influence(X) = \sum_{Y \in Followers(X)} \frac{1 + p \cdot Influence (Y)}{||Followers(Y)||}
	\end{equation}
Influence(X) is the estimated number of people who will read a tweet by person X. The probability that person Y retweets person X is assumed to be the constant factor p and the probability that Y reads a tweet by X is assumed to be $\frac{1}{||Followers(Y)||}$. The resulting score can be used as an indicator for X's influence and consequently for their CSC.

\subsubsection{PageRank with topical similarities.}
A similar analysis algorithm was presented by Weng et al. in 2010 \cite{Weng2010}. They introduced TwitterRank, which takes the topical similarity between users as well as the global link structure of the network into account. 
The global link structure is an aspect of PageRank and allows an interpretation of a user's influence similar to the authority of a website: user A's influence is high if the sum of influences on users is high, which themselves have a high influence on others.

The topical similarity goes beyond PageRank and TunkRank, as it allows the algorithm to regard someone's influence in different topics. This is of interest for the extraction of CSC, as it looks beyond a person's general popularity (which may originate from fame) and focuses on the content of the interactions.

The TwitterRank algorithm can be divided into the three main steps: topic distillation, topic-specific relationship network construction, and topic-sensitive user influence ranking. 
\begin{enumerate}
	\item \textit{Topic distillation} means that for each user, a profile with topics of interests is created. This can be achieved by analyzing the content of their tweets. A direct way to infer the topic is the use of hashtags that are already in the posts. However, Weng et al. observed a very low usage of hashtags in their data set \cite{Weng2010}. This is consistent with the research of Kywe et al., who analyzed 44 million tweets and found that hashtags were used in less than 8 percent of them \cite{Kywe2012}. Therefore, Weng et al. implemented Latent Dirichlet Allocation \cite{Blei2003} topic modeling to determine the most popular topics.

For each Twitter user a document is created that contains all their tweets. Topics are modeled as probability distributions over words. Each document is then modeled as a probability distribution over topics.

With other measures, such as similarity measures based on the Jensen-Shannon Divergence, it is then possible to calculate the topical difference between two Twitter users. \cite{Weng2010}
	\item Based on these findings, \textit{topic-specific relationship networks} are created under the assumption that Twitter users have different interests that are reflected in the identified topics. The resulting graph reflects the follower relationships along different topics. Twitter users are represented as vertices and edges are directed from followers to friends. \cite{Weng2010}

	\item In the last step the influence of users is measured with the \textit{TwitterRank algorithm}. First a Twitter user is chosen at random. Then a random walk is performed along the vertices of the graph. This random walk is topic-specific insofar that the transition probability between two Twitter users is dependent on their thematic overlap (as calculated in the first step). The transition matrix is defined by \cite{Weng2010} as follows:
	\begin{equation}
	P_t(i,j) = \frac{| \tau_i |}{\sum_{a: ~ s_i ~ follows ~ s_a} |\tau_a |} \cdot sim_t(i,j) \rm, ~ where
	\end{equation}
	$P_t(i,j)$ is the transition probability regarding topic $t$ from follower $s_i$ to friend $s_j$, \par
	$| \tau_{i} |$ is the number of tweets published by $s_{i}$, \par
	$\sum_{a: ~ s_i ~ follows ~ s_a} |\tau_a |$ sums over the tweets published by all of $s_i$'s friends, and\par
	$sim_t(i,j) $ is the similarity between $s_i$ and $s_j$ in topic $t$, which is defined as
	\begin{equation}
	sim_t(i,j) = 1 - |DT'_{it} ~ - ~ DT'_{jt}| \rm, ~ with
	\end{equation}
	$DT'$ being the matrix elements of the normalized $D \times T$ matrix, where D is the number of Twitter users and T the number of topics. A matrix element $DT_{it}$ of the unnormalized DT matrix represents the number of times a word in user $s_i$'s tweets has been assigned to topic $t$.
	
The transition probability captures two notions. It is dependent on the number of tweets by user $s_j$ that are theoretically visible to user $s_i$ because we can assume that $s_j$'s influence on $s_i$ increases together with her exposure to $s_j$'s tweets. Secondly, a topical similarity between the interests of both users is taken into account by the term $sim_t(i,j)$. \cite{Weng2010}

To address the problem of being stuck in a loop, which happens when a group of Twitter users follow only each other, Weng et al. use a transition vector $E_t$, with
	\begin{equation}
	E_t = DT''_{.t}.
	\end{equation}
$DT''_{.t}$ is the t-th column of matrix DT, which is the column-normalized form of matrix DT such that $||DT''_{.t}||_1$ = 1.
Thanks to this extension it is possible to jump to another random vertex instead of following the graph.	Now the TwitterRank $\bm{tr_t}$ of all Twitter users regarding topic $t$ can be defined with the iterative formula
\begin{equation}
\bm{tr_t} = \gamma P_t \times \bm{tr_t} + (1-\gamma ) E_t \rm, ~ with 
\end{equation}
$\gamma$ being a parameter between 0 and 1, which controls the probability of teleportation, and \par
$E_t$ being the transition vector along which the jump is carried out. \par
The result is a set of topic-specific vectors that contain a person's TwitterRank.	\cite{Weng2010} 
\end{enumerate}

To verify their results, Weng et al. worked with a data set of the top 1000 Singapore-based Twitter users that was obtained in 2009. On this data they analyzed the algorithm and identified the Twitter users with the highest TwitterRank in five categories. The fact that the number of followers of the top users varies widely is a consequence of TwitterRank's architecture that values not only the number of followers but especially the importance of the respective followers. Weng et al. also compare TwitterRank to the centrality measures In-degree (in this context the number of followers), PageRank (without topic sensitivity), and topic-sensitive PageRank. The Kendall correlation between TwitterRank and the other algorithms is 0.42 (In-degree), 0.47 (PageRank), and 0.68 (topic-sensitive PageRank). The fact that TwitterRank is most similar to topic-sensitive PageRank makes sense, because the other two measures are not directly topic sensitive. Additionally, the algorithms' performance in recommendation tasks was compared. Therefore, Weng et al. looked at randomly chosen following relationships between two Twitter users. Ten additional Twitter users without any relation to the follower were chosen. Then the link between the two original Twitter users was canceled. The task of the different algorithms was to identify the person among the other eleven who exerts the most influence on the follower. The performance of the algorithm is considered good if the original friend is chosen. This is then tested for several different scenarios and different chosen follower relationships. TwitterRank achieves the best recommendation quality in most scenarios, however it does not perform better in all cases. \cite{Weng2010} \par

TwitterRank and PageRank include information about the graph structure that goes beyond the pure count of performance indicators discussed by \cite{Anger2011}. There is no comparison to the ratios used by \cite{Anger2011}. However, the success of the algorithms, especially in the prediction task, show that there is relevant information in the graph structure that can be used for the prediction of a user's influence and thereby CSC. To predict concrete CSC values instead of performing a ranking, the TwitterRank of a person could be used as feature in a machine learning algorithm.

\subsubsection{Correlation of intrinsic metrics with a user's influence.}
In 2010 Cha et al. investigated several of the factors that we discussed in section \ref{se:intrinsicimplementation}: the number of followers, friends, tweets, retweets, and mentions \cite{cha2010}. Their research focus lies on the correlation of these factors with user's influence, i.e. the user's ``potential to lead others to engage in a certain act''\cite{cha2010}. As influence is a constituent of contributive social capital, the investigations can be used as input for the extraction of CSC values.

The data set collected by Cha et al. is comprehensive. It contains 55.0 million accounts that are connected by 2.0 billion social links and 1.8 billion tweets. After excluding inactive accounts (who published less than 10 tweets over the time of their existence) and private accounts (whose tweets can only be seen by friends), 6.2 million users were investigated.

Some of the discussed measures fail as indicators for someone's influence; however, others can give a good indication and corresponded to different aspects of a user's influence \cite{cha2010}:
\begin{itemize}
	\item A ranking based on the number of \textit{followers} correlates with the user's potential audience. Cha et al. conclude that people with a large In-degree are popular users: an analysis of the top 20 users shows that the accounts are primarily popular news sources, politicians, athletes, actors, and musicians. As the activity level of the followers is not considered, it does not necessarily mean that the user is influential in terms of spawning retweets and mentions. \cite{cha2010}
	\item	The number of times a user is \textit{retweeted} is seen as ``the ability of that user to generate content with pass-along value''\cite{cha2010}. An indication for this factual focus is that 92 percent of all retweets contain a URL. The top 20 users in this category are aggregation services, businesspeople, and news sites. \cite{cha2010}
	\item The number of \textit{mentions} a person receives can be seen as the user's ability ``to engage others in a conversation'' \cite{cha2010}. Less than a third of all tweets with mentions contain a URL. Therefore, Cha et al. interpret mentions as being less driven by the tweet's content and more by the author's identity. Therefore, they conclude that this measure reflects the user's name value. This is supported by the fact that most of the top 20 Twitter users in this category are celebrities \cite{cha2010}.
	\item The number of \textit{published tweets} as a sole measure for a user's influence often resulted in spammers or automated bots to be identified as most influential \cite{cha2010}. Therefore, this measure was disregarded.
	\item The \textit{outdegree}, namely the number of people a user follows (friends), suffers from the same problem and was consequently disregarded as well.
\end{itemize}

A normalization, e.g., dividing the number of retweets by the total number of published tweets, led to different results, that no longer ranked the users with the highest sheer number of retweets as influential. \cite{cha2010} 

Additionally, Cha et al. investigated the overlap between the three indicators followers, retweets, and mentions. As we have already seen from the brief discussion of the top 20 users, there may be only a small overlap. This is supported by an analysis of Spearman's rank correlation of the top ten percent of users, which confirms that the number of followers is not correlated to the other measures. Retweets and mentions correlate strongly (0.64), which shows that people who get retweeted often are also mentioned often, and vice versa. \cite{cha2010}

Cha et al. also analyzed the relation between influence and the discussed topics. For that the three most popular topics in 2009
were chosen. Tweets that discussed these topics were identified with several keywords that are associated with the respective topics. A period of 60 days was investigated. Longer time periods are often problematic, as spammers will try to use the respective popular keywords. The first finding is that there is only a small overlap of people who discussed all three topics -- only 2 percent of Twitter users who tweeted about one topic also mentioned the other two. The second finding is that influential users are in general not limited to only one topic but are influential over a variety of topics. \cite{cha2010}

Cha et al.'s findings are of high relevance for contributive social capital research on Twitter. The discussion of the metrics (the number of followers, tweets, retweets, mentions, and followees) and the corresponding types of influence is helpful for the definition of metrics and ratios for future social capital extraction. It is apparent that passive indicators (number of followers, retweets, mentions) by themselves are more important for the prediction of a user's influence than the metrics that can be directly altered by the user. Audience size (represented by followers), the ability to generate high quality content (retweets, mentions), and to engage in and socially maintain conversations (mentions, retweets) are properties expected of people with high CSC and value for their network. It was not investigated if a combination of these features would further improve the results. This was done by \cite{Hadgu2014} (see next subsection). The finding that a normalization of the features does not directly correlate with influence is opposed to the findings by Anger et al. \cite{Anger2011}, who were able to correlate normalized features to a user's influence. For CSC research this means that further research would be helpful, ideally with the help of CSC ground truth values.

\subsubsection{Expertise on Twitter.} A method that deals with the expertise of a person was presented in Hadgu and J\"aschke's 2014 paper \cite{Hadgu2014}. They investigate how various algorithms can be leveraged to identify researchers, who can be regarded as experts in their fields, on Twitter. Hadgu and J\"aschke used support vector machines (like Rao et al.), random forests, classification and regression trees, and logistic regression for the classification. The task of identifying researchers is one of binary classification, which is different from inferring CSC scores for each user.
We will therefore discuss the parts of their research that are of relevance for both tasks -- the used data set, the implemented methods and set the results in relation to CSC prediction. 

For the analysis, two data sets were collected and then combined. The first one consists of a random crawl of 1 million Twitter users. As the percentage of scientists among Twitter users can be assumed to be comparable to the average population, most of these users are not likely to be researchers. The second data set was selected with the intention of sampling a group of scientists. To achieve this task, the followers of the accounts of scientific conferences (e.g., @www09) were crawled. \cite{Hadgu2014}

To create a set with ground truth data (here: scientist yes/no), the users in the second set were matched against authors from DBLP, a computer science bibliography hosted at the University of Trier. The resulting list contains the positive examples. To create a list with negative examples (no scientists) the first data crawl was used. All users that follow any of the conferences and their followers were subtracted because of the increased likelihood of them being scientists. From the resulting list a random subset was drawn and used as ground truth set for people who are not scientists. \cite{Hadgu2014}

In the next step, Hadgu and J\"aschke defined features for the machine learning algorithms. Some features were obtained from profile information, like name, location, URL (e.g., if the top level domain is ``.edu''), description (e.g., if there are specific keywords like ``phd''), and total number of tweets, followers, and friends. Other features were selected from the content of the tweets, e.g., the number of tweets and retweets, different ratios (retweets to tweets, tweets containing URLs/hashtags), as well as distinct hashtags (e.g., conference names). The distinct hashtags were selected from the ground truth data set that only contains scientists. 1,872 hashtags were selected in that way. \cite{Hadgu2014}

With these features, various classification algorithms can be compared on the data set. Hadgu and J\"aschke performed stratified 10-fold cross-validation to train the models on 2,000 of the positive and all the negative profiles. The performance of all algorithms (support vector machines, random forest, classification and regression trees, and logistic regression) was very good. Random Forest performed best along all performance measures (precision: 0.96, recall: 0.92, F1-measure: 0.94, accuracy: 0.95, and true negative rate: 0.97). The other algorithms performed between 0.88 and 0.90 (precision), 0.87 and 0.90 (recall), 0.88 and 0.90 (F1-measure), 0.89 and 0.91 (accuracy), and 0.90 and 0.92 (TNR) and are therefore all useful for the analysis. \cite{Hadgu2014}

The importance of the different features was measured by their mean decrease accuracy, which shows how much using the feature in the classifier reduces the classification error. The most important feature was the number of tweets, followed by the number of tweets with scientific hashtags, the friend/follower ratio, and whether the user description contains keywords. \cite{Hadgu2014} \par

As stated in the beginning of this section, we are looking for methods to infer CSC values for each user. For this task we can leverage several findings from Hadgu and J\"aschke's research. For the CSC prediction, the machine learning classifiers need to be replaced with regressors and the ground truth values need to be replaced with CSC values. The used data set and the selected features can be used in the same way. The findings about the importance of the features can likely be transferred to the regression task, as the both tasks are similar in nature and use the expertise of a person as input. Consequently, the number of tweets, the number of tweets with scientific hashtags, and the friend/follower ratio should be used for CSC identification. Random forest, classification and regression trees, support vector machines, and logistic regression should all be used for regression analysis because they were successful for the classification.

\subsubsection{Summary: CSC on microblogging platforms.}
There is a variety of papers that deal with analyzing and categorizing users on the microblogging network Twitter. Some of the discussed procedures can be used directly for the extraction of contributive social capital, or its constituents influence and expertise. If a ground truth labeling with CSC values were available, one could use regression methods for the extraction of CSC from the presented data sets, as well as a mapping of the expertise and influence scores on the CSC per person. This would allow to give a structured CSC value that attributes parts of it to a user's influence as well as expertise. In the following, the key takeaways of the discussed investigations are summarized:
\begin{itemize}
	\item There are several useful features (number of tweets, number of followers, etc.) and ratios thereof (follower/following ratio, retweet and mention ratio, etc.) that can be either directly used to identify influential people, or as an input for analysis algorithms. Anger et al. showed, e.g., how these ratios relate to a person's influence \cite{Anger2011}, which is an attribute of social capital. A variety of additional features can be used as input for classification and regression algorithms (e.g., different linguistic features were proposed by Rao et al. and several others, like the use of parts of the user-profile, by Hadgu and J\"aschke).
	\item There are two main ways to crawl the data that is required for the analysis: random crawls and specific crawls of sub-networks to create data sets of people with specific properties (e.g., Hadgu and J\"aschke crawled scientists by selecting the followers of the accounts of scientific conferences to investigate expertise on Twitter). Both methods present advantages for CSC research. Random crawls extract a subset of the full network that shares the same characteristics and can therefore be used for the analysis. Specific crawls can be used as addition to create labeled ground truth data sets.   
	\item Ground truth labels on the data sets can either be obtained by manual labeling (e.g., to identify a person's regional origin based on language differences, see Rao et al.), or by selecting the data from Twitter lists (e.g., participant groups like the NRA, see Rao et al.). Additionally, one can match the user names with external databases (e.g., with scientific databases, see Hadgu and J\"aschke).
	\item Various algorithms can be used for the analysis: Intrinsic features (like the number of followers in Cha et al.), ratios (e.g., the Retweet and Mention Ratio by Anger et al.), PageRank-like algorithms (TwitterRank by Weng et al.), and supervised machine learning algorithms (support vector machines by Rao et al. and Hadgu and J\"aschke, decision trees and random forests by Hadgu and J\"aschke). All procedures could be used to identify CSC or its constituents on different microblogging datasets. Single features by themselves were less successful than methods that take more information about the network into account (see \cite{Weng2010}'s comparison of PageRank and TwitterRank vs. the count of followers).
\end{itemize}

\subsection{Social networking platforms} \label{socialnetworkingplatforms}
Social networking platforms are another data source that is relevant for the analysis of social capital.

Analogous to the procedure in the last section, we discuss different previous work that deals with the extraction of people's attributes, or the classification of people along different properties that are either directly relevant for social capital or indirectly, in the form of methods that can be adapted (e.g., with new ground truth data).

\subsubsection{Measuring influence with direct network features.}
In 2013, S. Hassan published a paper on identifying criteria for measuring influence of social media \cite{Hassan2013}. The main purpose of the study was to identify practical measures for influence in different social media communities. He investigated the intrinsic features as discussed in section \ref{se:intrinsicimplementation} and defined seven features as most relevant. Those were then categorized along three different dimensions of influence: 
\begin{itemize}
\item Recognition
\begin{itemize}
	\item Number of likes
	\item Number of subscribers/followers/friends
\end{itemize}
\item Activity generation
\begin{itemize}
	\item Number of posts
	\item Number of received comments on written posts
	\item Number of shares of the user's posts by others
	\item Number of in-links (number of times a user or her posts are referenced)
\end{itemize}
\item Novelty
\begin{itemize}
	\item Number of outlinks (how often resources outside of the website are linked)
\end{itemize}
\end{itemize}

These findings were based on a content analysis of social media. However, no quantitative assessment was used to verify the findings, e.g., by comparing against ground truth data. The identified criteria seem plausible and are in line with the previously discussed analysis on Twitter (compare for example Cha et al. \cite{cha2010}). This indicates a usability for social capital extraction of the named features, e.g., as input for machine learning algorithms.

\subsubsection{Formularizing social influence.}
In \cite{Bentwood2008}  J. Bentwood introduces the ``\textit{Social Media Index}.'' This index uses input from different types of social media to estimate the overall online influence of a person. The input is based on the six following criteria \cite{Bentwood2008}:
\begin{itemize}
	\item Number of friends on Facebook
	\item Google rank, in-bound links, subscribers, Alexa rank, content focus, frequency, number of comments of the person's blog
	\item Number of friends, followers, and updates on Twitter
	\item Number of contacts in LinkedIn
	\item Number of photos by and with the user on Flickr
	\item Favorites on Digg and del.icio.us
\end{itemize}

This multi-level approach is comparable to the Klout score, which is discussed in the next subsection.

Based on these findings, Bentwood at al. describe a person's influence/online presence in the form of an expression \cite{Bentwood2008}:
\begin{equation}
\rm \frac{Volume ~ and ~ Quality ~ of ~ Attention \times Time}{Size ~ and ~ Quality ~ of ~ Audience}
\end{equation}

The components of the expression are not clearly defined, which means there is no formal model of how to directly assess the quality of attention, for example. It can, however, be approximated with intrinsic features and tailored to the goal of the analysis. 

For the extraction of contributive social capital, the following metrics could be used as input for the expression. The volume and quality of attention could be measured with the number of friends a user has and the number of likes a user receives -- both metrics that were described by \cite{Hassan2013} as measures for recognition. An interesting part of the formula is the normalization in the denominator because it allows to take the audience's properties into account, which is related to the PageRank based approach by \cite{Weng2010}. By doing so, one can value the input of people proportionally to their respective contributive social capital. The size of the audience can be determined either by checking for the number of views of specific posts, or other metrics like the overall number of posts (to normalize for posts without reaction), the number of friends/followers (to show the ratio of responses), or the size of the network (to show the ability to reach people).


\subsubsection{Commercialized influence calculation with machine learning.} \label{kloutscore}
Rao et al. \cite{Rao2015} described an elegant approach to measure a person's online influence. they developed a machine learning algorithm that calculates the Klout score of a person. This score represents their influence on a logarithmic scale from 1 to 100. The used data sources are social networking platforms, like Facebook, Google+, LinkedIn, microblogging (Twitter and Instagram), and other web portals, like Wikipedia. Most of the other publications discussed in this paper use static data sets, which means the data was crawled once and then analyzed. Contrary to that the Klout score analyzes up to 45 billion interactions of up to 750 million users on a daily basis. In the following, we briefly describe the method, feature types, and validation mechanisms to make this commercial software comparable to other approaches. The algorithm works along several steps:
\begin{itemize}
	\item A feature vector is created for every user for each network or community that is used as data source.
	\item Each feature has a different weight for the computation of the overall influence score of the user.
	\item The feature weights were determined with a supervised learning model trained on labeled data. For the calculation, Rao et al. used non-negative least squares regression. The weights are not published.
	\item The result from the different data sources is hierarchically combined to create one score per user.
	\item In the last step the resulting value is scaled to a value between 1 and 100 on a logarithmic scale.
\end{itemize}

Users have to sign up to use the system. Once someone is registered, the required data can be extracted with different APIs based on the granted permissions. The Twitter data is extracted from Mention Stream\footnote{https://gnip.com/sources/Twitter/ (retrieved 2016-08-29)}. To obtain the ground truth data for the supervised learning algorithm, Rao et al. presented evaluators with pairs of people. The evaluators knew the people from their network. They labeled the person they found to be more influential. In order to prevent mistakes, all pairs were evaluated by several people. In total, over 1 million evaluations were used to determine the weights.

A total of 3,600 features can be used by the algorithm. Rao et al. do not publish a complete list; however, the following features are included:
\begin{itemize}
	\item Analysis of the interaction graph between users (graph structure and reaction types)
	\item Profile information (e.g., job title from LinkedIn)
	\item Number of friends on Facebook
	\item PageRank derived from Wikipedia
	\item Number of news articles that mention the user
\end{itemize}

Rao et al. used two different approaches to validate their findings. Both methods can be leveraged to design verification mechanisms for CSC extraction.

For the first experiment, which was conducted over the course of a year, various users were encouraged to write about certain perks they received. The reaction of their audience was measured. Plotting the average number of reactions to a post over the Klout score of the respective user, produced a monotonically increasing curve. This is a strong indication that there is a correlation between the Klout score and a user's ability to spread information. 

Another approach for validation is to study the overlap with other lists of influential people. In \cite{Rao2015}, Rao et al. publish two such comparisons between the Klout score and the ATP Tennis Player Ranking and the Forbes' list of most powerful women. The reasoning behind choosing both lists is not explained in \cite{Rao2015}. The overlap is determined with the normalized discounted cumulative gain metric (nDCG, see equation \ref{eq:ndcg}, \ref{eq:dcg}) and the calculated scores of 0.878 and 0.874 indicate a large overlap between both lists and the Klout ranking.

\begin{equation}
nDCG_p = \frac{DCG_p}{IDCG_p},
\label{eq:ndcg}
\end{equation}
where $p$ is a particular rank position and $IDCG_p$ is the ideal DCG through position $p$. 
\begin{equation}
DCG_p = \sum_{i=1}^{p}\frac{2^{rel_i}-1}{log_2(i+1)},
\label{eq:dcg}
\end{equation}
where $rel_i$ is the graded relevance at position $i$.

For the task of contributive social capital retrieval, one can leverage several of these findings. The definition of features (compare also S. Hassan \cite{Hassan2013} and J. Bentwood \cite{Bentwood2008}) that are then used in supervised learning is a method that can be extended to identify a person's contributive social capital with the help of different ground truths. Similarly, the data sources and the validation mechanisms can be adapted for CSC extraction.

\subsubsection{Network analysis with centrality measures.} 
An approach that is transparent and broadly usable is presented by J. Sun, who lists multiple graph analysis methods that are applicable for social network influence extraction in \cite{Sun2011a}. He discusses popular measures for social network analysis: degree centrality, closeness centrality, and betweenness centrality. These measures are often used for social influence analysis, because the resulting values are a sign for someone's importance within the network. The same argument can be used for contributive social capital analysis. A person whose centrality value is high is closely connected to other people within the network and is often close to or even the important node that connects different circles. This is exactly what is expected from people with high social capital. It also underlines the relevance of the measures used by J. Sun in the context of this paper.

In \cite{Sun2011a}, degree centrality is defined as 
\begin{equation}
	c_i^{\rm DEG} = deg(i).
\end{equation}
$c_i^{\rm DEG}$ is the number of nodes directly connected to node $i$. For an undirected graph, like the one created from the connections of a person on Facebook, this is equal to the number of friends an individual has. This was already discussed as a measure for a user's influence in \cite{Bentwood2008} and \cite{Hassan2013}. For directed graphs (e.g., the friend/follower structure of Twitter) one can separate between in-degree and out-degree (compare previous section on microblogging). It is furthermore possible to extend this method by looking at the k-path centrality. This measures the number of paths of length k that originate from node $i$ and thereby take the connectedness of $i$'s surrounding nodes into account. \cite{Sun2011a}

\textit{Closeness centrality} is not based on the volume of the originating paths but on the lengths thereof. To assess the importance of a node one calculates the average shortest path length to all other nodes in the network:
 \begin{equation}
	c_i^{\rm CLO} = e^T_i S\textbf{1},
\end{equation}
with $e_i$ being a column vector whose $i$th element is equal to 1 and all other elements are 0. S is a matrix whose ($i$, $j$)th element contains the length of the shortest path from node $i$ to $j$ and \textbf{1} is a vector whose elements are all 1. The lower the resulting $c_i^{\rm CLO}$, the shorter the average distance to the other nodes and the more central node $i$ is. \cite{Sun2011a}

Another measure that is popular in network analysis is the \textit{betweenness centrality}. It analyzes the position of a node within a network. Nodes of high betweenness are critical within the network, which happens, e.g., when they are at the interface between tightly-knit groups. Freeman's betweenness centrality is defined as follows
\begin{equation}
	c_i^{\rm BET} = \sum_{j,k}\frac{b_{jik}}{b_{jk}},
\end{equation}
with $b_{jk}$ being the count of shortest paths between the nodes $j$ and $k$, and $b_{jik}$ the number of times the shortest path goes through the investigated node $i$. Sun also discusses several variations that reduce the computation time of this approach. \cite{Sun2011a}

There is also a variety of other centrality measures, e.g., Feedback centralities (similar to PageRank, which was used for social media analysis in \cite{Weng2010}), Eigenvector centrality, as well as variations of the methods discussed above, and more \cite{Kosch2004}. 

Applying any of these centrality measures to a graph, extracted from a social networking platform like Facebook, yields aspects of the importance of a user within the investigated graph structure. As seen on Twitter, there is a correlation between a user's centrality measure and their influence \cite{Weng2010}. Depending on how the extracted graph was created this allows additional conclusions about the user's CSC. It is, e.g., possible to refine this methods in order to get a content- or domain-dependent score, which is achieved by reducing the network to edges between nodes that deal with certain topics (e.g., by only regarding discussions on computer science) and repeating the centrality analysis process. Additionally, the centrality measures can be used as features in supervised learning algorithms that predict CSC.

\subsubsection{Trust in online social networks.}
Another aspect of CSC was discussed by Ziegler at al. in \cite{Ziegler2009}. The focus of this work is the assessment of trust between agents from online social networks. For their investigations, they distinguish between local and global trust. Local trust is the personal feeling between different agents and the degree to which they trust each other. Global trust is the reputation of a person that can be inferred from information in the network.

Most of the algorithms used for the extraction of global trust work similarly to PageRank; the global trust of an agent is related to the global trust of the agents linking to him. Examples for such metrics are EigenTrust by Kamvar et al. \cite{Kamvar2003} and PowerTrust \cite{Zhou2007}.  Local trust metrics, on the other hand, ``take the agent for whom to compute trust as an additional input parameter and are able to operate on partial trust graph information.'' \cite{Ziegler2009} This allows for an individual trust assessment. In order to obtain a usable score, local trust metrics need to leverage structural information defined by personalized webs of trust. \cite{Ziegler2009}  \par

For CSC analysis, we are interested in an objective measure for the social capital an individual adds to the network and not primarily in the subjective assessment of parts of the network. Consequently, global trust metrics are of higher relevance. They are similar to PageRank, an algorithm whose importance for contributive social capital research was already underlined (compare section about TwitterRank \cite{Weng2010}).  The resulting lists can be used as input for CSC prediction.

\subsubsection{Summary: CSC on social networking platforms.}
As described in this section, there are various approaches that can be used in order to investigate social capital on social networking platforms. The discussed methods can either be used directly (if the investigated properties are constituents of CSC) or indirectly (by adapting the procedure, e.g., with the help of a new ground truth). The key takeaways of this section are summarized in the following:

\begin{itemize}
	\item Similar to microblogging, there are a number of intrinsic performance indicators that can be used as a measure for the extraction of CSC or its constituents. An overview of different metrics and their categorization was given by S. Hassan \cite{Hassan2013}. This is similar to the extraction methods used by Cha et al. for the microblogging service Twitter \cite{cha2010}.
	\item In order to tailor the extraction method to the specific interests of the study (e.g., social capital), these measures can be combined to get a better depiction of the reality. An exemplary formula was given by J. Bentwood in \cite{Bentwood2008}. This makes it possible to create ratios similar to the ones discussed by Anger et al. in the previous section (also see \cite{Anger2011}). We offered an interpretation of how this formula might look for CSC, based on the previous work by \cite{Hassan2013}. 
	\item The Klout score by Rao et al. (see \cite{Rao2015}) is an example for a machine learning algorithm with several different data sources. This measure uses supervised learning and was tested successfully in several experiments.
	\item As social networks can be represented as graphs, we discussed popular graph analysis methods: degree, closeness, and betweenness centrality (J. Sun \cite{Sun2011a}) and ways how they can be leveraged for CSC analysis. The resulting values can also be used as input for supervised learning algorithms.
\end{itemize}



\subsection{Direct communication} 
Whereas Twitter is limited to 280 characters per tweet and posts on social networking platforms are usually also kept short, there is no relevant limit for email communication. Additionally, emails are often used for the exchange of information. This makes the analysis of email networks so compelling for the extraction of CSC and its constituent expertise. The research already conducted in this field can be broadly categorized along two main targets: expert finding and social status/hierarchy extraction. Both topics show an overlap with social capital. Expert finding is connected to assessing the expertise of a person, which is a constituent of contributive social capital (see section \ref{su:overlap}). While the direct correlation between social capital and social status remains to be investigated, we still want to discuss it in this section because the social status of an individual can be regarded as consequence of this person's CSC. In the following we discuss several exemplary contributions to present the ongoing research by describing the method used, the data set and verification, as well as the findings.

\subsubsection{Expert identification using search algorithms.} 
In \cite{Zhang2005} Zhang et al. compare algorithms designed for expert identification. They evaluated searching strategies from three different families: general computational (breadth first search, random walk), network structure based (best connected, weak tie, strong tie, cosine similarity, hamming distance), and similarity based (information scent). These algorithms are used to create a match between a person with an information need and the best expert to provide the required information. The matching process itself is of little direct interest for social capital analysis. However, the discussion of the experimental verification method is relevant, as it can be leveraged in social capital extraction research.

The investigations were carried out with the help of the Enron data set, which contains about 500,000 emails from 147 employees. For the analysis a graph was constructed with 32,766 messages as edges between the 147 employee-nodes. An information profile for each user was created by using TF-IDF for all messages sent and received by the user. It is an approximation that these really represent the person's information space but there is no feasible way to get better profiles of the former Enron employees. 

For the evaluation Zhang et al. defined different queries and assigned them randomly to a node in the network. The eight search strategies were used to find a match between the `information seeker' and the person who was most likely able to provide the answer. The match was evaluated using a standard TF-IDF measure, therefore the person with the exact same combination of keywords within her information space would get the highest score. The highest success rate was achieved with breadth first search, closely followed by information scent, best connected, cosine similarity, and hamming distance.

Even though the focus of this work lies on the search algorithms, there are several aspects that can be leveraged for CSC research. By counting the number of times a person was identified as expert in a search, one can create a metric for each person's topic-sensitive expertise. As this metric would respond to how often the person helped others in the network, it should be  correlated to the person's CSC. This metric could also be used as input to supervised learning algorithms. Additionally, the data set is publicly available and can be investigated further.



\subsubsection{Expert identification with Bayes' theorem.}
Another paper that deals with the topic of expert identification based on email data was published by Balog et al. \cite{Balog2006}. They use Bayes' theorem to identify a person with the highest probability to be an expert according to a defined query. This allows to identify the top experts in different topics -- which could be used as input for CSC identification.

The probability that a candidate $ca$ is an expert in topic $q$ is given by \cite{Balog2006} with the help of Bayes' theorem as
\begin{equation}
p(ca|q) = \frac{p(q|ca)p(ca)}{p(q)}
\end{equation}
\cite{Balog2006} used the resulting probability $p(ca|q)$ to rank the experts for each topic $q$. For that, $p(ca|q)$ was calculated with two different models. Model 1 computes $p(ca|q)$ by constructing a candidate model from all documents associated with candidate $ca$ by collecting all term information from them. This candidate model is then used to represent the candidate in the query. The approach of model 2 is different. It assumes that $q$ and $ca$ are conditionally independent and that their relation can be described through document-candidate associations. With the help of these document-candidate associations the person is determined who is most strongly associated with the documents that describe the topic. This requires investigating each document (e.g., email) to identify who is associated with it. 

In both models, each candidate is scored by aggregating over all email documents associated with the candidate. The topical overlap between the emails and the query is calculated with language modeling. 

For experimental evaluation Balog et al. used the email W3C data\footnote{http://research.microsoft.com/en-us/um/people/nickcr/w3c-summary.html (retrieved: 2016-09-15)}, which comes with a list of candidate experts, topics, and relevance assessments for these topics. 
On this data, the association of a person to the email was investigated. The findings are summarized in the following. 

Model 2 generally outperforms model 1. Generally, there are four ways in which a person can be involved in an email: as sender, recipient, in cc, or within the message content. Balog et al. found that a person who is the sender of an email has the highest correlation to that person actually being the expert on the content of the email. \cite{Balog2006} Consequently, this feature should be included in future CSC investigation with the help of regression from features.

For CSC extraction the correlation of $p(ca|q)$ with CSC ground truth values can be investigated to verify to what extend $p(ca|q)$ can be used as input or proxy for the contributive social capital of a person.

\subsubsection{Expert identification with communication patterns and message content.}
In \cite{Campbell2003} Campbell et al. researched ways to identify experts in email communication data sets. Their objective was to identify experts using communication patterns as well as message content. 

The algorithm presented by Campbell et al. can be divided in three different steps. At first, all emails related to a topic are collected. This is achieved by keyword retrieval and clustering. In the second step a communication graph is created in which the nodes correspond to people and the edges are directed from email sender to recipient. In the third step a modified version of the HITS (Hyperlink-Induced Topic Search) graph-based ranking algorithm is implemented. 

Following \cite{Manning2009}, the HITS algorithm works as follows. For every query, each node in the network is assigned two scores: authority score and hub score. Originally these nodes were websites, in the context of CSC these nodes are network participants. People who are sources of information on the topic of the query are described as authorities, people who have a lot of connections to authorities, but are themselves no authorities, are called hubs. This leads to the circular definition that good authorities are linked to by many good hubs, and that good hubs link to many good authorities. These scores can then be computed in an iterative fashion.

With the reasoning that people send their questions to others who they think can give them an answer, Campbell et al. use the authority score as a measure for someone's expertise.

In order to evaluate this approach, Campbell et al. compare it to a query term frequency approach on two email data sets. Both algorithms were compared to a ground truth expertise score between 1 and 10. The query term frequency approach is similar to the one discussed by Balog et al. as it ranks a person according to the number of emails sent on a specific topic. The person ranked the highest in a specifictopic with this algorithm consequently sent the most emails related to this topic. One data set was extracted from a research organization (15 participants and 13,417 messages) and one from a software development organization (9 participants and 15,928 messages). The expertise score was created by averaging over ratings from other candidates. People who received an average expertise score of over 7 were marked as experts. \cite{Campbell2003}

The comparison was conducted with the help of several metrics: percent of correctly identified experts, percent of false alarms, precision, and detection (which combines correct identification, incorrect identification, incorrect rejection, and correct rejection). Overall the HITS algorithm did perform significantly better than the query-term frequency approach. The detection score of HITS was 0.39 (0.63) vs. 0.28 (0.44) for the research organization (software development organization). A similar advantage was observed for precision with 0.52 (0.67) for HITS vs. 0.38 (0.50) for the content-based approach. The percentage of false alarms was with 0.35 (0.55) vs. 0.71 (0.64) --- again significantly better for HITS. Only the percentage of correctly identified experts was better for the content-based approach, with 0.38 (0.33) vs. 0.44 (0.31) \cite{Campbell2003}. 

These findings are relevant for CSC extraction from email communication, as they underline that there is information about a person's CSC that can be extracted via analysis of the graph structure with the HITS algorithm in a way that is superior to straightforward term-frequency approaches. This also supports the findings of Weng et al. (\cite{Weng2010}) on Twitter (see section \ref{chap4twitter}) who successfully indentified influential users with other centrality measures. The authority scores calculated with the HITS algorithm can consequently be used as input for CSC retrieval and their correlation to CSC ground truth values investigated.

\subsubsection{Social hierarchy in email networks.}
The aspect of social hierarchy was investigated by Rowe et al. in \cite{Rowe2007}. They presented an algorithm to automatically extract social hierarchy structures from electronic communication networks. 

Like Zhang et al. they worked with the Enron email data set from 2002, which contains communications from employees along several hierarchies. The employees can be cross-referenced against the organizational charts.

The algorithm works as follows. For each user several features are analyzed to create a ranked user list. The first analyzed aspect is the information flow between users. The investigated features are:
\begin{itemize}
	\item number of sent and received emails (hypothesis: more important people need to communicate more),
	\item average response time (hypothesis: the faster user $i$ answers to user $j$'s request, the more important $j$ is to $i$.)
\end{itemize}
The next aspect Rowe et al. investigated is the communication network graph that was created in a way that all users are vertices that are connected by undirected edges if they exchanged at least N emails. In this context the following features were calculated for each user:
\begin{itemize}
	\item number of cliques an account is associated with,
	\item raw clique score that also takes the size of the cliques into account,
	\item weighted clique score that regards the importance of other clique members (determined by the number of emails and average response time),
	\item centrality measures: degree-, closeness-, and betweenness-centrality as well as the clustering coefficient,
	\item hubs-and-authorities importance of each user.
\end{itemize}

These individual metrics are then weighted and a social score between 1 and 100 is computed along which the users are ranked. \cite{Rowe2007} 

The results were compared with the organizational chart and show a great overlap, as Rowe et al. were able to predict the social hierarchy of the 54 traders in the North American West Power Traders division of Enron with great accuracy. A higher standing within an economic setting usually translates to better access to resources. Therefore, these findings can be used for the information access (expertise) aspect of CSC and the correlation of the social score and ground truth CSC values investigated. 


\subsubsection{Social status in email communication.}
In \cite{Bird2006} Bird et al. investigated the extraction of social status of email users. As data set they used the email list of an open source software (OSS) development project. Over the course of seven years Bird et al. collected roughly 100,000 messages of more than 1000 users.

To evaluate a person's importance within the network Bird et al. used the centrality measures betweenness, in-degree, and out-degree. According to these measures, the developers among the OSS group are higher in status than the non-developers, which is reasonable given the nature of the data set.
Within the group of developers there is also a significant correlation between the centrality measures and the number of committed changes to the code. This indicates that the developer's importance correlates to these measures. Therefore, one can assume that the centrality measures are relevant for the CSC extraction from email communication. This underlines the findings by \cite{Campbell2003} and \cite{Weng2010} that information about the constituents of CSC of users is stored in the graph structure of communication networks, which can be accessed via centrality measures.   

\subsubsection{Summary: CSC on email communication networks.}
The key findings of this section can be summarized as follows.
\begin{itemize}
	\item There are less features available than on microblogging or social networking platforms, as, e.g., there is nothing similar to the ``likes'' on Facebook. Features that are relevant for the CSC of users are the number of emails sent and received, response time, clique measures, and centrality measures (as presented by Rowe et al. \cite{Rowe2007} and Bird et al. \cite{Bird2006}). This is similar to the previous findings in other data sources.
	\item The success of centrality measures (e.g., betweenness centrality, HITS) for the prediction of CSC constituents shows that there is relevant information within the graph structure of the investigated networks (see \cite{Campbell2003}, \cite{Rowe2007}, \cite{Bird2006})
	\item For expert finding the usual approach is to compute an overlap between a defined query and the information space of a user. This can be done with TF-IDF (see Zhang et al. \cite{Zhang2005}) or with other language models (compare Balog et al. \cite{Balog2006}).
	\item The standard data sets used for the investigation of email communication stems from companies or research organizations, e.g., the Enron email corpus.
\end{itemize}

These findings are of relevance for the expertise aspect of social capital, as the content-focus of email communication make it ideal for expert identification as well as social status and hierarchy assessment. The presented approaches and algorithms can again be used directly for the extraction of the CSC constituent expertise, or indirectly by adapting them for CSC extraction (e.g., using the discussed features and centralities as input for supervised learning algorithms with CSC as ground truth values). There are many similarities between email communication and other means of direct communication, e.g., similar features (sender, recipient, length of the message, message content, etc.) and  graph structure.  Due to these similarities, it is likely that the findings can be extended to cover these other networks (e.g., WhatsApp communication), as well. When doing so, it is important to adjust for the differences, e.g., by adding new features (groups in WhatsApp) or removing them (CC recipients in emails is not available in all means of direct communication).

\subsection{Citation Networks} 
As discussed in section \ref{su:definitions}, a key aspect of someone's CSC is their knowledge and expertise. By providing an expert's opinion on various topics, people can help their surrounding network making decisions and building opinions. This increases their value within the network. A type of networks that directly deal with expertise are scientific citation networks. In section \ref{se:intrinsicimplementation} the nature of citation networks were briefly presented and citations were mentioned as direct performance indicators. However, there is a number of methods that go beyond the pure count of citations. In the following publications about citation network analysis are discussed and the findings are set in relation to contributive social capital extraction.

\subsubsection{Hirsch index, g index and i10 index.}
There is a variety of indices that are used throughout the scientific world to measure a scientist's importance.

A popular index was introduced by J. Hirsch and sets the number of publications of an author in relation to the number of times they are cited: ``the index h is defined as the number of papers with citation number $>$h, as a useful index to characterize the scientific output of a researcher''\cite{Hirsch2005}. This makes the resulting score robust against outliers, e.g., scientists who publish very many papers with very little positive feedback (e.g., the count of citations), or scientists who find one golden nugget which gets cited very often but don't publish anything else. J.Hirsch verified the validity of the h-index on a data set of physicists and found that scientists with a high h-score are indeed prominent. He concludes that h gives an estimate of their ``importance, significance, and broad impact'' \cite{Hirsch2005}.

Abbasi et al. investigated the Spearman correlation between an author's social capital and their h-index and citation count \cite{Abbasi2014}. They found a positive correlation of .57 (h-index) and .44 (citation count). This shows that the h-index is a better measure to approximate someone's social capital than the pure count of citations. \cite{Abbasi2014}

With the g-index, L. Egghe presented a variation of the h-index in \cite{Egghe2006}. It is a refinement of the h-index insofar that it also looks at the number of publications of an author and the respective citations. Additionally, it gives credit to very successful articles, which is important to identify the most successful experts. The value of g is defined in a way that the top g articles of a scientist were cited (altogether) at least $g^2$ times.

Google Scholar introduced the i10-index, a straightforward measure that shows the number of articles with at least ten citations\footnote{http://googlescholar.blogspot.de/2011/11/google-scholar-citations-open-to-all.html (retrieved 2016-07-04)}. By ignoring the publications that promoted few responses, this metric focuses on popular articles. However, this also means that lots of publications that receive only low levels of attention are disregarded. The same happens with new publications that are still rising in popularity but have not passed the threshold yet. 

These indices can be used as a measure for the expertise aspect of contributive social capital, optionally after categorizing along different topics. They can also be used as features for machine learning algorithms.

The indices could also be implemented for CSC investigations on collaborative question and answering portals or microblogging platforms. For that they need to be adjusted (e.g., the number of tweets that received at least h retweets). 

\subsubsection{Trust and reputation with PageRank and graph analysis.}
In \cite{JAsang2007} J\o{}sang et al. recognize that the standard metric ``count of citations'' is an indicator of quality and reputation, however they also see shortcomings, like the fact that only positive referrals are possible. In order to counter this issue, they suggest the investigation of scientific citation networks with a PageRank like algorithm. Others later realized this, among them Yang et al. \cite{Yang2010}. 

In 2010 Yang et al. implemented several analysis methods for citation networks, among them PageRank in \cite{Yang2010}. The goal of their studies is closely related to identifying the CSC of a scientist. Based on scientific network data Yang et al. propose a method to estimate a researcher's importance, contribution, and reputation as well as a ranking according to these findings. We discuss the methods they investigated, the data sources used for evaluation, and their results.

The focus is the investigation of three methods that implement properties of both, topical link analyses as well as citation graph analyses:
\begin{itemize}
	\item Multi-type citation graph analysis (including relationships of authors, papers, affiliations and publishing venues) combined with content-based analysis,
	\item Heterogeneous PageRank model,
	\item Topical PageRank model.
\end{itemize}
For the multi-type citation network analysis, graphs were created with the nodes being either authors, papers, affiliations, venues, or combinations thereof. The (directed) edges between those social actors are all types of possible relations, like co-authorships, citations, publications in venues. \par
Heterogeneous PageRank could be directly applied to these multi-type citation networks by evenly distributing a node's authority among the connected nodes (authors, papers, affiliations, or venues). However, Yang et al. reason that this approach does not correctly represent the directional probabilities between the different types of social actors. Therefore, they propose the following approach, which assigns different propagation probabilities to the different types of out links \cite{Yang2010}:
\begin{equation} \label{eq:pagerankcittion}
PR(i) = (1 - d) \sum_{j:j\rightarrow i} \beta_{ji} \frac{PR(j)}{O(j)_{type(i)}} + d \frac{1}{N} \rm, where
\end{equation}
\begin{itemize}
	\item $j$ is the node pointing to node $i$,
	\item $d$ is a random jump,
	\item $\beta_{ji}$ represents the propagation probability from node $j$ to $i$ and is always identical for jumps to nodes of the same type. Also $\sum_{type(i)} \beta_{ji} = 1$ holds true, which means that the probability to jump to nodes of one of the different types is always normalized to 1. 
	\item $O(j)_{type(i)}$ is the number of outlinks from node $j$ to nodes of the same type as $i$.
	\item $N$ is the total number of notes of the network.
\end{itemize}

The next step in their analysis is the introduction of a topical element by introducing different probabilities for the random jump. This allows following a link within the network, jumps to a random link within the same topic, or random jumps to another node in a random topic. This takes the notion into account that an author's expertise in one subject is not directly transferable to other topics.

For experimental verification, Yang et al. used a crawl of the ACM digital library\footnote{http://dl.acm.org/dl.cfm (retrieved 2016-09-09)} which consists of 172,891 distinct web pages that represent different publications (including publishing venue, authors, affiliation of each author, and citation references). This data was then used to create the aforementioned graphs. 

The ground truth data, i.e. rankings of authors in different topics, was needed for experimental comparison of the different methods. This data was obtained in three ways.
\begin{itemize}
	\item Authors were labeled with the help of the Libra rankings of conferences they attended.
	\item The ACM database labels were used (ACM fellow, ACM distinguished, ACM expert).
	\item Four judges reviewed the Google Scholar and other search results and labeled the scientists based on the web pages found.
\end{itemize}
Normalized discounted cumulative gain (nDCG, see equations \ref{eq:ndcg}, \ref{eq:dcg}) was then used as a metric to compare the three different methods for queries in 23 categories. The main results can be summarized as follows \cite{Yang2010}:
\begin{itemize}
	\item The multi-type citation graph (with nodes being authors, papers, venues) leads to better results than graphs that only use one of the types as nodes. Therefore, the affiliation between authors and the venues, where the papers are published, also provide useful information. This finding can be expanded to the other data sources as well. On social networking or microblogging platforms, e.g., a multi-type graph could include the user's posts or tweets as nodes, additionally to the users themselves.
	\item Using the topical PageRank improves the performance.
	\item Homogenous PageRank can be improved by applying parameter tuning.
\end{itemize}
These methods can be leveraged to extract the knowledge aspect of social capital of scientists. The determined rank can be used as input for CSC analysis.


\subsubsection{Expert identification in communities.}
In 2012, Su et al. published \cite{Su2012}, an article about identifying the top k experts within a community. They put special focus on the topic context in order to capture all aspects of an author's expertise. Additionally, it was investigated how these search results can be diversified. This task by itself is of little relevance to CSC retrieval, but the methods, the data set, and the results can be leveraged to assess the knowledge aspect of social capital in social networks.

The first step in Su et al.'s method is to identify topic information. They discuss several methods, from using pre-defined categories, via user assigned tags, to statistical topic modeling. The chosen method is the author-conference-topic (ACT) topic model that was presented by Tang et al. in 2008 \cite{Tang2008a}. This model allows to calculate the probability of a connection between an author or paper to certain topics.

Additionally, Su et al. defined an objective function that evaluates the precision with which the algorithm ranks the experts vs. ground truth data. The performance is influenced by the weights that are given to the features which are publications, h-index of the author, and the language model-based relevance score. The weights are determined by maximizing the objective function with the help of a standard greedy optimization algorithm. After the weights are set, the algorithm can be used to determine the top k experts for each query. 

The data set used for evaluation is an unnamed scientific network with more than 1 million authors and 2 million papers. In order to create the ground truth subset, a similar approach to the one presented by Yang et al. was used. For each investigated search query the best conferences were chosen. Then the chairs and committee members of these conferences were ranked according to appearance and h-index. The resulting top 100 people in the list were labeled experts and chosen as ground truth.

Su et al. compare their algorithm against several alternative methods using language model, topic model, and random walk and found that it improves the ranking performance.

For CSC retrieval both contributions can be leveraged; the topic-model to determine the topical context of a user's expertise, and the ranking using an objective function and an effective algorithm to solve it. The resulting ranks can be used as input to determine a user's contributive social capital.

\subsubsection{Citation network analysis with Mendeley metrics.}
In 2013, Li et al. investigated the social media platform Mendeley\footnote{https://www.mendeley.com/ (retrieved 2016-09-02)} that gives users the chance to archive, comment, and share their paper bibliographies \cite{Li2013}. This yields several new features that can be used for the analysis, for example the total number of readers of an article. The focus of Li et al. was to identify influential scholars along two dimensions: academic and social impact. In this section the data set, methodology, and findings are briefly presented.

The data set was extracted from Mendeley in 2012 and comprises approximately 1 million user profiles, 100,000 papers, and various types of connections between the users, like co-authorship, contacts, or group memberships. 

The academic impact of a scholar is high if her publications are popular and received well by peers. The assumption of Li et al. is that the most relevant criteria for this is the number of times a paper is read. Considering that a lot of people read papers and do not necessarily cite them, this is a reasonable assumption. Based on the number of readers they define three different metrics\cite{Li2013}:
\begin{itemize}
	\item Total number of readers (indicates the author's overall influence)
	\item Maximum number of readers per paper (to identify scholars with few but very visible papers)
	\item R-index (an R-index of n represents that the author has published n papers with at least n readers. This is a measure of a scholar's productivity and impact and similar to the h-index for citations.)
\end{itemize}

Li et al. plotted these three metrics over the respective number of scholars and identified power law patterns. This indicates that there are only a few scientists that can be regarded as real influences, whereas the majority have very low scores. Another observation is that all three measures yield different scientists in the top one, respectively top ten percentile\footnote{The overlap values for the top 10\% (top 1\%) of scientists for the different measures are: R-index vs max reader count: 24.4\% (7.9\%), R-index vs total reader count: 45.5\% (29.8\%), max reader count vs total reader count 77.7\% (62.5\%).}. Therefore, the three measures can be used to describe different types of influential scholars. \cite{Li2013}

For the analysis of the social influence of an author Li et al. were interested in a user's connections within the network and their ability to control information flow. They used the three centrality measures that we discussed in the context of social networking platforms: degree-, closeness-, and betweenness-centrality (compare Sun et al. \cite{Sun2011a}). Li et al. used scientists with at least one published paper as nodes and all interrelations between them (e.g., co-authorships) as edges. The result were several sub-graphs, as there is little to no overlap between different scientific fields. The centrality measures were then used to investigate local research communities. These were created using a Latent Dirichlet Allocation \cite{Blei2003} topic model to generate topic vectors for each paper and cosine similarity to compute the similarity between the topic vectors of the papers and of 25 characteristic search queries. Afterwards, for each node a single influence value was created by calculating the Euclidian norm of the normalized scores of the three measures. The scholars identified in this way are mainly young researchers --- opposed to those identified to have high academic influence, who tend to be older. Li et al. also used Spearman's Rank Correlation Coefficient to investigate the overlap between academic and social influence and found no linear correlation. \cite{Li2013}\par
Both measures can be used as input to investigate different aspects of social capital. The academic influence stems from scientific expertise and highly recognized papers and is therefore relevant for the expertise and knowledge aspect. Social influence is a measure for a person's connections and engagement within the community. Therefore, one should be able to use a combination of both aspects for the assessment of a scientist's CSC. This could be achieved by averaging over both influence values and then assigning a CSC score, or by using the metrics (number of readers, R-index, centrality measures, etc.) as input for supervised machine learning (similar to \cite{Su2012}).

\subsubsection{Centrality measures in scientific networks.}
Another paper that is dedicated to the analysis of scientific networks was published by Kas et al. in 2012. In \cite{Kas2012} they investigate the change of scientific networks over time and in this context also discuss measures that are of relevance for social capital analysis. 

In order to identify the key authors in the field of high-energy physics, Kas et al. used centrality measures similar to those discussed in the previous section (compare \cite{Li2013} and \cite{Sun2011a}). They argue that degree centrality can be used to identify intelligence, closeness centrality to a source to transmit/acquire information, and betweenness centrality as a measure to identify a person who connects groups. 

For the analysis Kas et al. used a publicly available data set compiled by arXiv with 29,555 papers in the field of high-energy physics and the citation graph between them. 

Citation and betweenness centrality were applied only on the co-authorship network. Degree centrality was also implemented in the form of publication out-degree, co-authorship degree, and citation in-degree. The networks created in this sense are comparable to the multi-type citation networks discussed by Yang et al. in \cite{Yang2010}. 

There is no comparison to ground-truth data but the overlap between the publication out-degree and co-authorship degree is quite high, which is in line with other investigations in the field \cite{Kas2012}. The citation in-degree is interpreted as the prestige of an author and Kas et al. do not observe a significant overlap between this ranking and the other centralities. \cite{Kas2012}

For the extraction of CSC these differences indicate that a single centrality measure may not be sufficient to assign a CSC score. However, they can be used as input for further investigations.


\subsubsection{Summary: CSC in citation networks.}
The key takeaways of this section in the context of social capital retrieval, are:

\begin{itemize}
  \item There are several indices (h-index, r-index, i10 index) that can be directly used to assess a person's scientific influence. \cite{Hirsch2005} \cite{Egghe2006} The h-index correlates with a user's social capital \cite{Abbasi2014}.
	\item There is a variety of different features (maximum number of readers, publications, citations, R-Index, h-index) that can be used as input for algorithms to determine someone's expertise (see Li et al. \cite{Li2013}, Su et al. \cite{Su2012}).
	\item Different types of PageRank (homogenous or topical) can be implemented to determine a participant's influence (see Yang et al. \cite{Yang2010}).
	\item Other centrality measures can also be used to assess a person's influence within the network (Li et al. \cite{Li2013}, Kas et al. \cite{Kas2012}).
\end{itemize}

The extracted expertise, influence, and importance of a person, which is often in the form of a ranking of the scientists, can be used as input to determine CSC scores. If one is interested in dividing social capital into different topics, one can use topic models (e.g., those done by Su et al.) or leverage the fact that scientists usually publish in well defined fields with little overlap \cite{Li2013}.


\subsection{Threaded discussion boards} 
Threaded discussion boards are among the most popular websites in the world and are consequently interesting for CSC assessments from online data sources. Along with previous work about the two discussion boards Reddit and Slashdot, we also discuss publications about online question and answer portals. Those are generally organized along various categories and allow users to discuss problems in a similar fashion.

\subsubsection{Social network analysis of Reddit.}
The scientific research about Reddit so far was mainly focused on the existing voting mechanism, which was described in section \ref{se:intrinsicimplementation}. E. Gilbert, e.g., shows one limitation of the current mechanism in \cite{Gilbert2013}. Even though it is the purpose of Reddit to promote new and interesting ideas, E. Gilbert demonstrated that more than 50 percent of links that later turned out to be popular, were overlooked on their first submission. A. Richterich makes another interesting observation: ``Karma functions as main, quasi-monetary incentive and reward of participation''\cite{Richterich2013}. Even though this reward has no value in the sense of real currency, many participants strive to maximize their points, e.g., by stealing content and claiming it as their own. \cite{Richterich2013}

The fact that Reddit is a large, openly available collection of topic-centered discussions makes it well suited for further investigations. A straightforward way to improve the existing metrics could be to
\begin{itemize}
\item separate all comments into different domains (topic distillation), and
\item normalize the obtained Karma per comment (Karma/Comment ratio).
\end{itemize}

Considering the number of views per page (similar to the method presented by \cite{Gilbert2013}), the number of responses to a comment, or the number of reposts and references a popular comment creates is another way to create more features for contributive social capital analysis. These features can then be used in machine learning algorithms to predict CSC values. 

\subsubsection{Social network analysis on Slashdot.}
In the book ``Computing with Social Trust'', J. Golbeck reviews previous work on trust in online communities \cite{Golb2009}. As there are millions of users who post daily on Slashdot.org, the authors conclude that ``it is often a good strategy to delegate the quality assessment task to the users themselves'' \cite{Golb2009}. This is achieved by the intrinsic ranking scheme where users can award others with 'karma', similar to Reddit's approach. For the assessment of social capital this has the same consequences: an algorithm that extracts CSC scores can leverage these intrinsic karma scores.


To assess the applicability of other network analysis metrics for threaded discussion boards one can leverage the findings of G\'{o}mez et al., who analyzed the social network and discussion threads in Slashdot in \cite{Gomez2008}.

For their investigations G\'{o}mez et al. created a graph based on the relations between authors of posts and the people who reply. They discarded several posts (e.g., anonymous ones or posts that did not inspire a discussion) and ended up with a network of more than 80,000 people and almost 1.3 million comments. \par
Their key findings that can facilitate future social capital research are \cite{Gomez2008}:
\begin{itemize}
	\item The Slashdot network shows features of traditional social networks, for example large components (disconnected sub-graphs within the network), small average path length, and high clustering. 
	\item The differences between Slashdot and traditional networks are a lower reciprocity (users do not reply as often as observed in other social networks) and neutral assortative mixing by degree (highly connected users do not prefer the exchange with other highly connected users which is often observed in other networks)
	\item Connections between users are less explicit as in other social networks, where they can be made public as ``friendships''. 
	\item Two classes of users exist that can be separated by the mean score of their posts. Good writers achieve on average higher scores and run-of-the-mill writers lower scores. 
\end{itemize}
The similarities between the Slashdot network and other social networks suggest that similar analysis methods can be implemented. The fact that different user classes exist additionally supports the idea of underlying CSC values that vary from user to user.

\subsubsection{Trust and reputation on Slashdot.}
In their 2009 paper Skopik et al. focus on trust and reputation in professional virtual communities, which include discussion forums like Slashdot. In \cite{Skopik2009} they present a system which determines trust relationships between community members automatically and objectively by mining communication data. This is relevant for the trust component of CSC and addresses the problem that human based ratings can be unfair and biased.

The algorithm assigns a value to comments that attract other comments, depending on the number of answers a post inspires. This algorithm is then extended to model trust by implementing two functions. The first function represents the confidence of user $i$ in user $j$. Skopik et al. set this function equal to the discussion strength between both users in a specific topic. The next function models the reliability. This function is set to 1 if user $i$ interacts with user $j$ at least 10 times within a year on a specific topic. The trust between both users is then calculated as a product between confidence and reliability function. \cite{Skopik2009}

The experimental validation was conducted on data that was crawled from two Slashdot subdomains over the course of one and a half years. After the data were cleaned (e.g., subtracting anonymous users) the graph had 24.824 nodes and 343.669 edges. The sets of trusted users identified by the algorithm were similar to the sets identified by humans. \cite{Skopik2009}

The calculated trust can be used as input for CSC extraction. It is also possible to refine the results, for example, by implementing more complex functions for confidence and reliability. 

\subsubsection{Authority identification in Q\&A portals.}
In their 2015 paper, Bouguessa et al. propose methods to identify authoritative actors in online communities, like question and answer portals. As question and answer portals are content focused, it is reasonable to assume that good contributors achieve a higher authority score than others. Consequently, their contributive social capital should be higher -- which underlines the relevance for this paper. 

Bouguessa et al.'s approach can be divided into two steps. At first, a feature vector is derived for each user that contains information on the user's social activities. Then, a statistical framework, based on the multivariate beta mixtures, is used to model the estimated set of feature vectors. This is used to identify the most authoritative users in the network. \cite{Bouguessa2015} 

In the question and answer platform Stack Exchange, Bouguessa et al. used the following characteristics as input.
\begin{itemize}
	\item Number of answers,
	\item Number of answers that were chosen as ``best answer'' by the community,
	\item Number of votes received,
	\item Z-score $= \frac{a-q}{\sqrt{a+q}}$, where $a$ ($q$) is number of answers (questions) by the user. 
\end{itemize}
To best distinguish between regular and important users, these features are transformed to a logarithmic scale. This compresses the long tail of higher values that correspond to authoritative users, while stretching out the smaller region which corresponds to a lot of less influential users (e.g., low count of best answers). Additionally, all features are normalized into the interval $[0,1]$. 

Bouguessa et al. propose that these normalized user feature vectors follow a mixture density. Its parameters can be determined with maximum likelihood estimation. Authoritative users can then automatically be determined by choosing people whose feature values lie above a certain threshold. 

A verification of this algorithm on Stack Exchange and Twitter data yielded high quality results \cite{Bouguessa2015}. 

The features, as well as the multivariate approach, can be investigated for CSC extraction.

\subsubsection{Summary: CSC in threaded discussion.}
The most relevant findings for CSC extraction from threaded discussion boards and question and answer portals are summarized as follows.
\begin{itemize}
	\item Richterich and Gilbert discussed the current rating algorithms of Reddit \cite{Richterich2013} \cite{Gilbert2013}. Based on their findings we concluded that the intrinsic evaluation tool (``Karma points'') should be investigated for social capital extraction and propose to implement ratios for normalization as we have seen for Twitter (compare Anger et al. \cite{Anger2011}).
	\item Similarly, for the online community Slashdot, J. Golbeck suggests the use of the internal quality assessment mechanisms to evaluate user's trust \cite{Golb2009} --- a constituent of CSC. 
	\item There are similarities between Slashdot and other social networks (see work by G\'{o}mez et al. \cite{Gomez2008}) that suggest that similar analysis methods can be used to investigate user's features and consequently social capital.
	\item The problem of human bias can be countered by automated trust assessments (Skopik et al. \cite{Skopik2009}).
	\item The features on question and answer portals can be used to identify authoritative users (Bouguessa et al. \cite{Bouguessa2015}).
\end{itemize}




\section{Summary and conclusions}\label{se:summary}
At first, we discussed the term (contributive) social capital and identified several of its aspects that were already the target of research in previous work. There are four aspects that we mainly focused on in this paper: reputation, trust, influence, and expertise. Reputation is the general belief of the public towards a person. Trust is a more individual sentiment which is often founded on personal experiences between users. Influence is, among others, the ability to change other people's way of thought. Expertise is a measure related to knowledge and experience of an individual in a defined topic. All of these terms are important parts of a person's online presence and measuring them can help to clarify and improve all types of online interactions. 

We selected five main fields of online interactions: microblogging, social networking platforms, scientometrics, direct communication, and threaded discussion boards. Most of these sources have built-in measures that indicate a person's performance on the respective platform. On Twitter the most popular metrics are followers, retweets, and mentions. On the social networking platform Facebook the number of friends, likes, received and posted messages can be used as indicators. In direct communication the popular intrinsic metrics are the number of conversation partners as well as the count of received emails. In scientometrics we have citations, and threaded discussion boards often implement up- and down-vote mechanisms. 

Additionally to these straightforward metrics, there is a number of methods that can be employed to obtain information on specific attributes of the participants.

In \textbf{microblogging} one can form ratios with the direct features (Follower/ Following Ratio, Retweet and Mention Ratio, etc.) and then set the result in relation to aspects of CSC, like influence (compare Anger et al. \cite{Anger2011}). This is a straightforward extension to the use of single features (see Cha et al. \cite{cha2010}). With the help of classification algorithms, one can classify users according to different aspects. Linguistic details (Rao et al. \cite{Rao2015}) or additional user information (Hadgu and J\"aschke Hadgu2014) can be used for this. 

A variety of algorithms has been investigated. Weng et al. introduced TwitterRank, which is similar to PageRank and produces good results in this area \cite{Weng2010}. Additionally, supervised machine learning algorithms (support vector machines by Rao et al. and Hadgu and J\"aschke, and Random Forest, Classification and Regression Trees, and Logistic Regression by Hadgu and J\"aschke) can be used for the extraction ofCSC values. 

On \textbf{social networking platforms} there are also publications that investigate the relation of the intrinsic features with aspects of CSC (see S. Hassan \cite{Hassan2013}). These features can be set in relation to build ratios (e.g., the exemplary formula given by J. Bentwood in \cite{Bentwood2008}). Supervised machine learning was used by Rao et al. in \cite{Rao2015} to assess a person's influence. Another way to determine a participant's standing within her network is the use of centrality measures. Degree-, closeness-, and betweenness-centrality measures (see \cite{Sun2011a}) can be leveraged for social capital analysis, which was shown successfully in citation networks (\cite{Li2013}), direct communication (\cite{Campbell2003}, \cite{Rowe2007}, \cite{Bird2006}), and microblogging (\cite{Weng2010}).

For \textbf{direct communication} we have seen several ways to collect data sets for analysis. Examples are the Enron company data, the W3C email data, or collections of open source projects. The aspect of social capital that is usually researched in this context is expertise, because of the content-focus of email conversations. An approach to identifying experts is to calculate the overlap between keywords of a predefined query and the information space of a user. This can be done with TF-IDF (see Zhang et al. \cite{Zhang2005}) or with other language models (compare Balog et al. \cite{Balog2006}). Features that are relevant for the classification of users are number of emails, response time, clique measures, and centrality measures (as presented by Rowe et al. \cite{Rowe2007} and Bird et al. \cite{Bird2006}). Campbell et al. used a HITS algorithm on an email graph and identified experts correctly \cite{Campbell2003}. 

\textbf{Scientific citation networks} are another data source that is relevant for social capital extraction. A range of indices can be directly used to assess someone's scientific influence: h-index, r-index, and i10 index \cite{Hirsch2005} \cite{Egghe2006}. One can also create features by leveraging additional information (e.g., the maximum number of readers) and combine this with indices. Li et al. \cite{Li2013} and Su et al. \cite{Su2012} used this as input to determine a person's expertise. Different types of the centrality measure PageRank (homogenous or topical) can be used to determine a participant's influence in scientific networks (see Yang et al. \cite{Yang2010}). Li et al. \cite{Li2013} and Kas et al. \cite{Kas2012} computed the influence of network participants with other centrality measures. Another effective approach is to use algorithms that determine the weights of relevant features by optimizing an objective function on ground-truth data (Su et al. \cite{Su2012}). 

Richterich and Gilbert discussed the current rating algorithms of the \textbf{threaded discussion board} Reddit \cite{Richterich2013} \cite{Gilbert2013}. Based on their findings we concluded that the intrinsic evaluation tool (``Karma points'') should be leveraged for social capital extraction and propose to implement ratios for normalization. For the online community Slashdot, J. Golbeck also suggests the use of internal quality assessment mechanisms to evaluate user's trust \cite{Golb2009}. Additionally, there are similarities between Slashdot and other social networks (see, e.g., the work by G\'{o}mez et al. \cite{Gomez2008}) that suggest that similar analysis methods can be used. In \cite{Skopik2009} Skopik et al. proposed methods for automated trust assessments in order to address the bias of human voting. 

There are commonalities between all five data sources. Previous work shows that contributive social capital or some of its constituents can be extracted from all of them. As described in detail above, this can be achieved by using intrinsic features, ratios thereof or by leveraging centrality measures or machine learning algorithms. As all these methods produced some kind of correlation or indication for the investigated user's CSC, we conclude that a successful analysis should implement some or all of these procedures. This could be achieved by using supervised machine learning with CSC values as ground truth and the discussed metrics and centrality values as features.

This work points to tools, data sets, verification mechanisms, and ideas to conduct further research on the topic of contributive social capital extraction to test this hypothesis.






\end{document}